# Perfect compensation of absorption in metamaterials for diffraction-unlimited imaging

**W. Adams[1], and D. Ö. Güney[1*]**

[1]Department of Electrical & Computer Engineering, Michigan Technological University, 1400 Townsend Dr., Houghton, MI 49931-1295, USA

[*]dguney@mtu.edu

**Abstract**-To overcome the resolution limit of conventional optics, near field imaging techniques using a negative index flat lens (NIFL) have been previously developed that amplify the evanescent components of the incident field. Here, a technique is developed and demonstrated to compensate for losses in a non-ideal NIFL by determining the transfer function of the lens and subsequently applying the inverse to the unresolved raw image. The result is a compensated image with sub-diffraction-limited resolution.

In conventional imaging systems, the field components with spatial frequency greater than $\omega/c$, where $\omega$ is angular frequency of the wave and $c$ is the speed of light in the medium, become evanescent and decay rapidly. Typically, the image sensor is located far from the source such that all the evanescent components are decayed beyond the sensitivity of the sensor. Therefore, the largest spatial frequency components a conventional imaging system can detect is theoretically limited to $\omega/c$; the so-called diffraction limit. In order to capture higher spatial frequency components and increase the resolution of imaging systems beyond this limit, Pendry proposed a near field imaging technique with a slab of negative index material [1]. In this work, a new near field imaging technique is presented using a lossy negative index flat lens (NIFL) in conjunction with the post-processing loss compensation technique inspired by plasmon injection scheme [2] and described in Fig. 1. The process begins with an image produced by a NIFL. A compensation filter is then applied to the image in order to amplify the high spatial frequency components which have been attenuated. The compensation filter is the inverse of the transfer function of the NIFL, which can be obtained analytically or from numerical simulation.

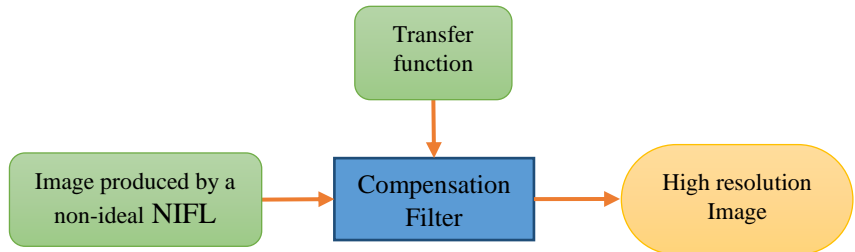

**Figure 1** Block diagram of the proposed near field imaging technique.

Unlike the previous near field imaging methods, this technique does not benefit from a lossless NIFL. The technique developed here is based on a non-ideal NIFL with a practically achievable figure-of-merit (FOM) for loss, defined by the ratio of the real and imaginary parts of the refractive index [3-5].

The optical properties of the NIFL are defined in terms of complex relative permittivity $\varepsilon_r = -1 + j\varepsilon''$ and complex relative permeability $\mu_r = -1 + j\mu''$ with $\varepsilon'' = \mu'' = 0.1$, resulting in a FOM of about 10 and an impedance match with free space. In this work, the transfer function for the NIFL was obtained by numerical simulation in COMSOL. The incident wavelength $\lambda_0$ was chosen to be 1μm, and the thickness $2d$ of the NIFL was chosen to be 0.5μm. $d$ was chosen as thin enough to minimize the loss experienced by high spatial frequency components of the incident field while still providing sufficient amplification, in addition to considering practical concerns such as the unit cell size of possible constitutive negative index metamaterials, mechanical strength of the lens, and challenges with optical alignment of submicron imaging systems since the thickness of the lens determines the distance from the object and image planes.

In order to perform imaging with the NIFL, a simulation was configured with an arbitrary "object" electric field defined at the object plane with sub-diffraction feature size ($\lambda_0/4$). A raw image was then recorded at the image plane that is subsequently compensated with the technique in Fig. 1. The result is a sub-diffraction compensated image as seen in Fig. 2. It can clearly be seen that while the raw image does not provide a resolved image, the compensated image results in a near-perfect reconstruction of the object.

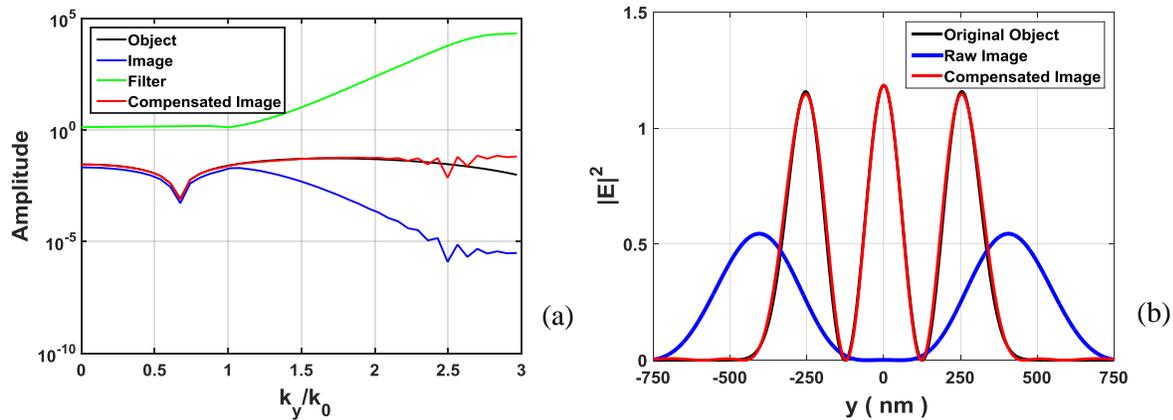

**Figure 2** (a) Fourier transform of the object and image demonstrating compensation of high $k_y$ attenuation and (b) calculated intensity of the object, raw image, and compensated image. ($\lambda_0 = 1\mu m$)

Previously, near field imaging techniques have been proposed using a NIFL to achieve image resolution beyond the diffraction limit. However, compensation of losses in a non-ideal NIFL has been largely ignored. Here, the transfer function of a lossy NIFL is determined by simulation, and the inverse is used as a filter to recover lost high spatial frequency components of the raw image that is produced. In Fig. 2(b), it can be seen that the resulting raw image from the lossy NIFL does not resolve the original object containing sub-diffraction feature size. After application of the compensation filter in Fig. 2(a), many of the lost high spatial frequency components are recovered, resulting in the compensated image shown in Fig. 2(b). The compensated image displays sub-diffraction-limited resolution.

This work was supported by Office of Naval Research (award N00014-15-1-2684) and the National Science Foundation under Grant No. ECCS-1202443.